\documentclass[runningheads]{llncs}
\usepackage{blindtext,graphicx}
\usepackage[absolute]{textpos}
\usepackage[n,
advantage,
operators,
sets,
adversary,
landau,
probability,
notions,
logic, ff, mm, primitives, events, complexity, asymptotics, keys,
]{cryptocode}
\usepackage{enumitem}
\setlistdepth{20}
\renewlist{itemize}{itemize}{20}
\setlist[itemize]{label=}

\usepackage{etoolbox}
\makeatletter
\preto{\@verbatim}{\topsep=2pt \partopsep=2pt }
\makeatother

\usepackage{fancyhdr}

\usepackage{array}
\usepackage{algorithm}
\usepackage{algorithmic}
\usepackage{attrib}
\usepackage{mathtools}
\usepackage{hyperref}
\begin{document}

\title{Risk Framework for Bitcoin Custody Operation with the Revault Protocol}

\author{Jacob Swambo\inst{1,2} \and
Antoine Poinsot\inst{2}}

\authorrunning{J. Swambo  \& A. Poinsot}
\institute{King's College London, Department of Informatics \\
\email{jacob.swambo@kcl.ac.uk} \and
WizardSardine \\
\email{darosior@protonmail.com}}

\maketitle              
\thispagestyle{fancy}

\begin{abstract}

Our contributions with this paper are twofold. First, we elucidate the methodological requirements for a risk framework of custodial operations and argue for the value of this type of risk model as complementary with cryptographic and blockchain security models. Second, we present a risk model in the form of a library of attack-trees for Revault -- an open-source custody protocol. The model can be used by organisations as a risk quantification framework for a thorough security analysis in their specific deployment context. Our work exemplifies an approach that can be used independent of which custody protocol is being considered, including complex protocols with multiple stakeholders and active defence infrastructure.

\end{abstract}

\section{Introduction}

While mainstream acceptance of Bitcoin as an asset appears to be increasing, advanced tools and methods for secure custody of bitcoins are slow to develop. Bitcoin custody encompasses the protection of assets through software, hardware, and operational processes. The foundation of Bitcoin custody is key-management, a well understood topic in the academic literature and in practice. However, Bitcoin custody, in particular multi-stakeholder custody, involves human processes, communication protocols, network monitoring and response systems, software, hardware and physical security environments. Given a secure cryptographic layer, there are still vulnerabilities introduced at the application layer by software developers, at the hardware layer throughout the supply chain, and at the operations layer by users. Without adequate risk management frameworks for custodial operations, Bitcoin users are likely to suffer unexpected losses whether they self-custody funds or employ a third-party custodian.  

Open-source custody protocols are emerging \cite{revault-pdf,Glacier,Subzero,Swambo2020vault} and are a critical ecosystem component for improving security standards. If a custody protocol stands to public scrutiny and offers a high-level of security without relying on proprietary processes, users, insurance companies and regulators can have more confidence in it. The emerging custody protocols are trying to reconcile the needs of traditional businesses and banking  with Bitcoin's novel identity-less and irreversible transaction properties. A lack of available and accepted open-source custody protocols means that organisations are heavily relying on third-party custodians, or deploying their own custody protocol.

We propose an attack modelling technique as the basis for a risk framework for Bitcoin custody operations, using the Revault protocol\footnote{Specifically, the version identified as 609b40dda07155abe5cd4a5af77fc2211d11fbc1 which can be found on the open-source repository hosted on Github \cite{practical-revault}.} as a case-study \cite{practical-revault,revault-pdf}. While the process of model construction is intensive, the resultant framework is extensible and modular and some of its components can be re-used with different custody protocols. It is intended to be readily comprehensible, and, given sufficient validation, the framework can be used by any organisation intending to deploy Revault to better understand their risk posture. 

Risk quantification frameworks address several ecosystem problems. Organisations that control bitcoins or other digital assets need accurate models to engage in realistic risk-management.
The complexity of custodial risks leaves insurance companies guessing rather than systematically estimating when pricing their insurance offerings or assessing particular solutions for digital custody. 
Finally, emerging regulatory standards for custody \cite{DACS,CryptoassetCustody} are simple and fail to capture advanced custody architectures or enable context-specific risk analyses that acknowledge the full security environment of a custody operation.    

The remainder of this paper is structured as follows.
Section \ref{sec:RevaultOverview} summarises the components and processes of the Revault protocol. Section \ref{sec:Methodology} discusses our evaluation criteria for an operational risk framework, and introduces the attack-tree formalism on which our risk model is based. Section \ref{sec:Threat Model} presents our operational risk model for Revault. Section \ref{sec:Conclusion} concludes this paper. 

\section{Overview of Revault Custody Protocol}
\label{sec:RevaultOverview}

Revault is a multi-party custody protocol that distinguishes between \textit{stakeholders} and fund \textit{managers}.
The primary protection for funds is a high-threshold multi-signature Script controlled by the stakeholders. 
The day-to-day operational overhead of fund management is simplified by enabling portions of funds to be delegated to fund managers.  Stakeholders define spending policies in-line with traditional controls of expenses, and have automated servers to enforce their policies. In addition, a deterrent is withheld by each stakeholder to mitigate incentives to physically threaten the stakeholders. To achieve this, Revault makes use of sets of pre-signed transactions coupled with an active defence mechanism for detecting and responding to attempted theft transactions. In the following, we will outline the components of the Revault architecture, the transaction set,
 the stakeholders' routine signing process and the managers' spend process. Refer to \cite{practical-revault} for the detailed specification of the open-source protocol.

\subsection{Revault Architecture Components}
\label{subsec:Components}

Each stakeholder and manager has a \textit{hardware security module} (HSM) to manage their private keys and generate signatures for transactions. A backup of private keys is stored for each HSM in a separate protected physical environment. 

Each stakeholder and manager uses a \textit{wallet} software to track their co-owned bitcoins, craft transactions, store transaction signatures and communicate with each other through a \textit{coordinator}. The coordinator is a proxy server that simplifies communication for the multi-party wallet. All communication uses Noise KK encrypted and authenticated channels \cite{NoiseProtocolFramework}.

Stakeholders each have one or more \textit{watchtower}, an online server that enforces the stakeholder's spending policy limitations. 
Stakeholders each have an \textit{anti-replay oracle} server.

\subsection{Revault Transaction Set}
\label{sec:TransactionSet}

\begin{figure}[t]
\centering
    \includegraphics[width=12cm]{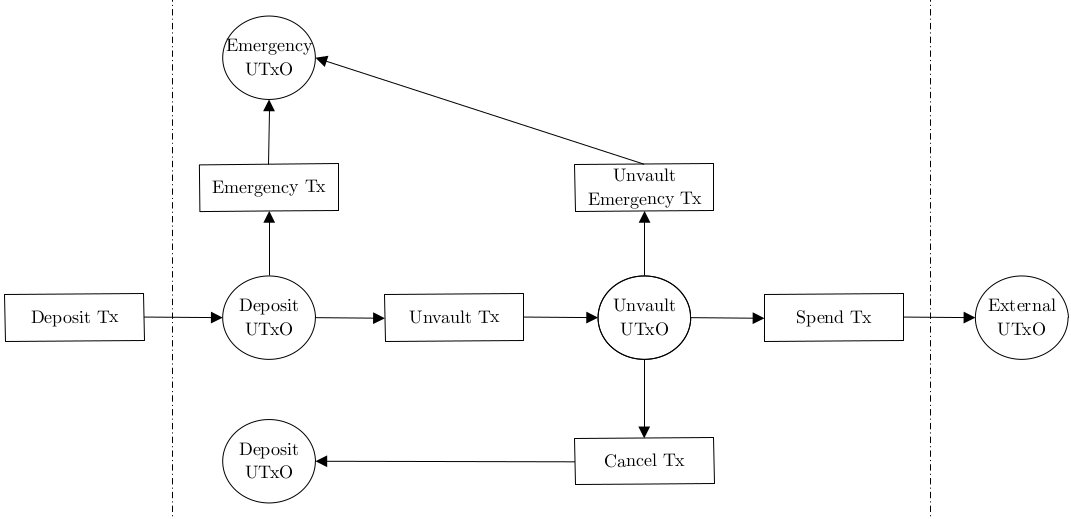}
    \caption{Diagram of the transaction (Tx) set structure in the Revault protocol. An Unspent-transaction-output (UTxO) is created by a preceding Tx and is consumed by an input in a proceeding Tx.}
    \label{fig:tx_diagram}
\end{figure}

The use of hierarchical deterministic wallets means that each participant in the Revault protocol has a tree of public and private keys \cite{BIP32}. To discuss ownership of bitcoins, we refer to a generalisation of a locking Script, called a \textit{descriptor}. The wallet will have multiple addresses that correspond to a single abstracted descriptor. Funds are deposited into the multi-party wallet through a Deposit transaction (Tx) output that pays to the deposit descriptor, describing $N-$signatories locking Scripts derived from the stakeholders' (\textit{stk}) extended public keys (\textit{xpub}). In descriptors language formalisation \cite{ScriptDescriptors} it is defined  as: \begin{verbatim}
    thresh(N, stk_1_xpub, stk_2_xpub, ..., stk_N_xpub)
\end{verbatim}

The set of transactions prepared with stakeholders' wallets and signed using their hardware security modules (HSM) include the Emergency Tx, Unvault Tx, Unvault-Emergency Tx and Cancel Tx. The managers can only prepare and sign a Spend Tx type. Figure \ref{fig:tx_diagram} depicts these transactions and the essential unspent-transaction-outputs (UTxOs) they create or consume.

An Emergency Tx locks funds to an emergency descriptor which is unspecified by the Revault protocol and is kept private among stakeholders. The descriptor must however be harder to unlock than the deposit descriptor. This is the deterrent for physical threats to the stakeholders. 

An Unvault Tx consumes the deposit UTxO and creates an unvault UTxO locked to the unvault descriptor, \begin{verbatim}
    or( thresh(N, stk_1_xpub, stk_2_xpub, ..., stk_N_xpub), 
        and( thresh(K, man_1_xpub, ..., man_M_xpub), 
             and( thresh(N, oracle_1_xpub, ..., oracle_N_xpub),
                  older(X) ) ) ),
\end{verbatim}
that is redeemable by either the $N$ stakeholders \textit{or} the $M$ managers (\textit{man}) along with $N$ automated anti-replay oracles after $X$ blocks.

A Cancel Tx consumes the unvault UTxO and creates a new deposit UTxO. The watchtowers' role is to broadcast the Cancel Tx if a fraudulent spend attempt is detected (either through an unauthorised attempt at broadcasting an Unvault Tx or if a Spend Tx does not abide by the spending policy).  The time-lock gives watchtowers $X$ blocks worth of time to broadcast a Cancel Tx. 
An Unvault-Emergency Tx consumes the unvault UTxO and locks funds to the emergency descriptor. It has the same purpose as the Emergency Tx, only it consumes the unvault UTxO rather than the deposit UTxO.
A Spend Tx is used by managers to pay to external addresses.

\subsection{Stakeholders' Signing Routine}

\begin{figure}[t]
\centering
    \includegraphics[width=9cm]{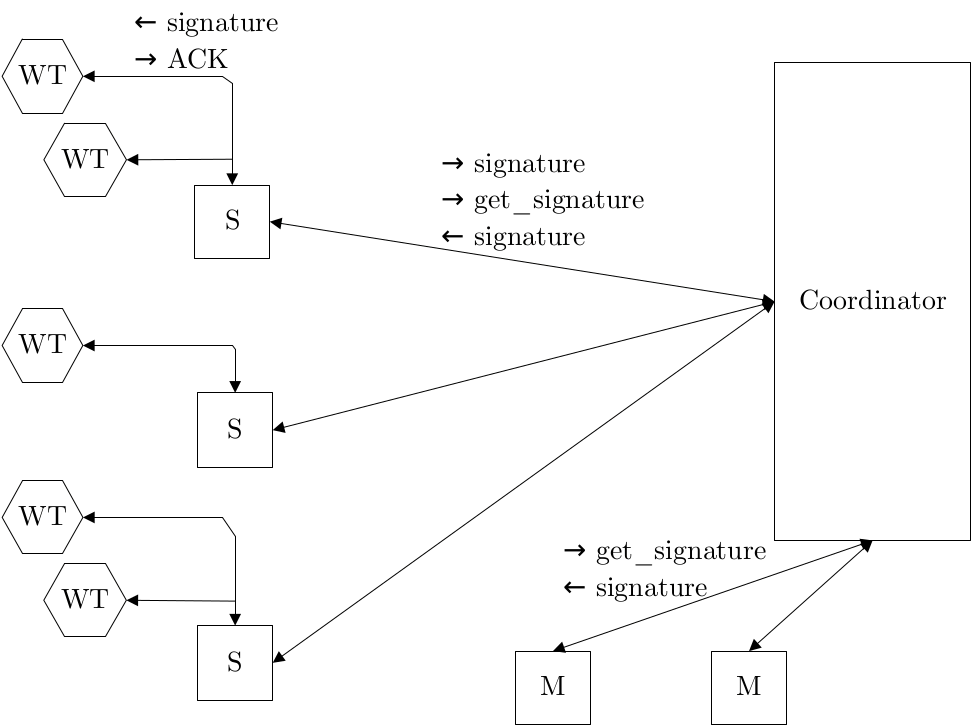}
    \caption{Data flow diagram for the communication of the \textit{stakeholders' signing routine} with an example Revault deployment. There are three stakeholders (S) who each have one or two watchtowers (WT). There are two managers (M) and a coordinator. Signature messages, signature requests and watchtower acknowledgements (ACK) are only shown once per connection type but apply to each connection of that type (e.g. there is \{$\leftarrow$ signature, $\rightarrow$ ACK\} between each WT and S).}
    \label{fig:sig_exchange_diagram}
\end{figure}

Stakeholders' wallets routinely check for new deposits 
and each one triggers a signing routine. Figure \ref{fig:sig_exchange_diagram} shows the connections and message types for an example Revault deployment enacting the signing routine. The wallet crafts an Emergency Tx and requests the stakeholder to sign it using their HSM. The stakeholder will verify the emergency descriptor on the HSM before authorising the signature generation\footnote{This feature is not available with current HSMs, but integrating compatibility with descriptors (along with other security features) would improve the human-verification component of HSM security and is being discussed on the \textit{bitcoin-dev} mailing list \cite{AdvancedHMfeatures}.}. 
The wallet then connects to the coordinator to push its signature and will fetch other stakeholders' signatures.

Optionally, stakeholders may also sign the Cancel, Unvault-Emergency, and Unvault Txs to securely delegate funds to the managers. In this case the signing process is the same but is carried out in two steps: first, the signatures for the Cancel and Unvault-Emergency Txs are exchanged with the other stakeholders through the coordinator and then shared with the watchtower(s), and only then are the Unvault Tx signatures shared with managers.

\subsection{Managers' Spending Process}

Most spending policies cannot be inferred from the Unvault Tx alone and so the Spend Tx must be known to the watchtower to validate an unvaulting attempt. In these cases the Spend Tx must be advertised to the watchtowers before unvaulting, otherwise it will be cancelled. The anti-replay oracle is required to avoid the Spend Tx being modified by the managers \textit{after} the unvault time-lock expires and thus by-passing enforcement of the watchtowers' spending policies.

\begin{figure}[t]
\centering
    \includegraphics[width=9cm]{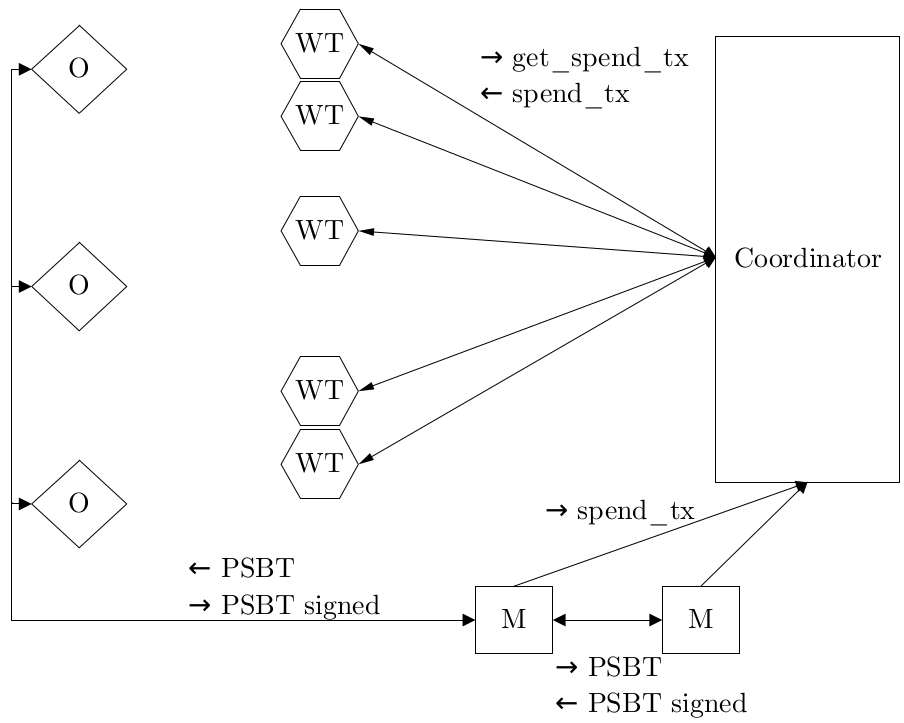}
    \caption{Data flow diagram for the communication of the managers' spend process. In this example there are two managers (M), three anti-replay oracles (O), five watchtowers (WT) and a coordinator. A Partially-signed Bitcoin Tx (PSBT) is exchanged among managers and between a manager and the anti-replay oracles. A fully signed SpendTx is shared with the WTs through the coordinator.}
    \label{fig:spend_tx_exchange}
\end{figure}

Any manager can initiate a spend. Figure \ref{fig:spend_tx_exchange} depicts the spend process. The initiator creates a Spend Tx, verifies and signs it using their HSM and passes it back to the wallet in the partially-signed Bitcoin Tx (PSBT) format \cite{BIP174}. It's  exchanged with a sufficient threshold of the other managers to add their signatures and hand it back to the initiator. The initiator requests a signature from each of the anti-replay oracles and pushes the fully-signed Spend Tx to the coordinator.  
The initiator broadcasts the Unvault Tx, triggering a lookup from the watchtowers to the coordinator for the Spend Tx. If the Spend Tx is valid according to \textit{all} of the watchtowers policies and none of them cancel this unvaulting attempt, the manager waits $X$ blocks and broadcasts the Spend Tx. 

If, during the unvaulting process, there's a significant increase in the fee-level required for a Spend Tx to be mined, a manager needs to bump the fee. Managers use a dedicated single-party fee wallet for this. Similarly, watchtowers use a fee wallet in the case there is high demand for block space to bump the fee for Cancel or emergency Txs. 

\section{Methodology}
\label{sec:Methodology}

To see where this research fits in to the big picture, consider the key life-cycle of a custodial operation. There are three phases; initialization, operation, and termination. Initialization is where wallet and communication keys are generated, where software integrity is verified, hardware security modules are checked, and relevant public information is shared among participants. Operation encompasses the active fund management. Termination is the phase wherein the wallet is de-commissioned and all sensitive information destroyed. Initialization and termination are out of scope for this paper. Our risk model covers the operations phase. In the following we present our rationale for our chosen attack modeling formalism and explain how this can be used as a risk framework.

\subsection{Operational Security Models}

A framework for high-level risk analyses for the integration of custody into a multi-stakeholder context has not yet been presented. To-date the literature has focused primarily on cryptographic security modeling, dealing with low-level risks associated with cryptographic primitives, key-management protocols, HSMs and single-party wallets. The underlying cryptographic security is fundamental but should be complemented by an operational security model, which is much more likely to be the domain where participants create vulnerabilities for an attacker. Advanced custody protocols that use multi-layer access control with both static and active defences for insider and external attackers demand a whole-system approach to security analysis.

We present now several requirements for our modelling formalism: a) the ability to represent complex processes with numerous components and sequential events; b) supports qualitative risk analysis; c) supports automated quantitative methods for multi-attribute risk analysis; d) readily comprehensible and visual models that are more amenable to open-source intelligence; and e) extensible and modular models to support differential analysis and re-use of modules.  

The two most popular attack modeling techniques in cyber-security literature are attack-trees and attack graphs \cite{AGraphATreeReview}. In short, tools for attack graphs tend to produce graphs that aren't readily comprehensible due to the complexity of real-world attack scenarios \cite{SurveyAttackModeling}. That is, attack graphs don't scale well \cite{AssetCentricAnalysis}. On the other hand, attack-trees seem to meet all of our requirements, at least when considered with the right structure and semantics (as described in section \ref{sec:Attack-treeFormalism}) and thus we construct our risk model using this formalism.

While a statement such as `our custody solution is based on an $m-$of$-n$ security model' can entail a lot for simple multi-signature custody protocols, it doesn't capture the reality nearly as well as our proposed methodology would. It is certainly not sufficient for a more complex custody protocol like Revault. What is the physical environment for those $n$ private keys? Are any of those keys online? Are there key backups and, if so, what protections are in place for these? Too much depends on the broader security environment of a custody protocol for it to be left without scrutiny.   

Application threat modelling has been used to harden the Revault protocol throughout both its theoretical development and implementation. For each application process (spend, routine signing, emergency, revault) a component-by-component and connection-by-connection analysis has been carried out to determine the consequences of outages, data tampering, component corruption, etc., and has resulted in the design specification \cite{practical-revault} and the transaction flow threat model \cite{revault-pdf}. The application threat modeling approach is complementary and has informed us in enumerating the risks presented with the attack-trees. However, in contrast to attack-trees, it lacks a semantic structure which is amenable for automated risk quantification and thus isn't suitable as the basis of a risk framework.

\subsection{Attack-Tree Formalism}
\label{sec:Attack-treeFormalism}

The risk model is presented using the formalism of attack-trees \cite{AttackTrees,ThreatLogicTrees,ThreatTrees}. Attack-trees have an attack at their root, and branches that capture alternate (OR) and complementary (AND) attack pathways comprised of intermediate attack goals as non-leaf nodes and basic attack steps as leaf nodes. As in numerous other works
\cite{ADTool2,DBLP:journals/corr/JhawarKMRT15,AFTrees,10.1093/cybsec/tyaa020,Duqu2,ADTreesSAND}, we extend the basic attack-tree to support  sequential conjunction of branches (SAND) allowing us to model an attack where some sub-tree of an attack pathway has to occur before and in addition to another sub-tree. For brevity we depict our attack-trees as nested lists. The logical gates (OR, AND, SAND) shown with each node apply to the next node at the same depth. This means that at any given depth, a node with a SAND gate occurs \textit{before} other nodes that are shown below it. Some aspects of the system are built to be resilient to attack and failure through redundancy. For example, an attacker needs to compromise all stakeholders' private keys to steal funds locked to the deposit descriptor. To be concise, rather than having several copies of the same sub-tree we write ($X$ times) to note that the sub-tree has to happen $X$ times. During an analysis, these sub-trees should be considered as $X$ separate AND sub-trees, since they are contextually different (corresponding to different participants, remote and physical environments).

We provide a set of attack-trees, capturing prominent risks that have been enumerated primarily by considering tangible and intangible assets. Tangible assets (bitcoins) are distinguished by the access control structures determined by the set of descriptors. We consider operational privacy and business continuity as intangible assets. 

Our work here is focused on security, rather than safety. In principle, the same methodology could be extended to an integrated security-safety model by constructing attack-fault-trees \cite{AFTrees}. Another common extension to the attack-tree formalism is to include countermeasures, producing attack-defence-trees \cite{FoundationsADTrees,ADTreeValue}. The benefit of our modular modelling technique is that it enables future work to integrate these extensions and re-use results from this work. Hence, we prioritise constructing a strong foundation based only on attacks, and aim to incrementally improve on the model presented as new intelligence emerges.

\subsection{On Risk Analysis}

Our purpose in constructing the risk model presented in section \ref{sec:Threat Model} is to provide a framework to support both qualitative and quantitative risk analyses for specific deployment instances of Revault custody. By determining costs, likelihoods, and other attributes for risks associated with custodial processes, an organisation can perform a differential analysis of countermeasures until their risk-tolerance is satisfied. An explicit framework not only helps an organisation deploying Revault with risk-management but could form a standard by which insurance companies and regulators consider specific deployments. As with any model of complex reality, attack-trees are imperfect and cannot capture every possible attack pathway, but the alternative---complete ignorance---is not better. 

To perform a context-specific risk analysis, a set of estimates are made (using in-house empirical data, public research, and expert opinion) for each basic attack step on different attributes such as monetary cost, execution time, or likelihood. With that, a bottom-up procedure (from leaf nodes to the root) is used to compute aggregated attributes. Bayesian methods can be used to update prior estimates with more refined values as new data sources emerge. The process for generating estimates is critical and should be considered with care. In-depth research-based practical guidance on this topic is given by D. W. Hubbard and R. Seiersen in \cite{Hubbard}.
Given specific contextual information, estimations can be improved by further decomposing basic attack steps (e.g. `steal keys backup') into multiple steps (e.g.  `bribe manager to determine backup location' SAND `break into safe'). If a basic attack step has a highly uncertain estimate, then further decomposition into more explicit steps can be beneficial. On the other hand, decomposing into quantities that are more speculative than the first could compound uncertainty rather than reducing it. 

Various methods for analysis can be used to compute aggregated attributes for attack-trees. Kordy \textit{et al.} gave an overview \cite{KordyReview}. Our purpose here is to provide the framework on which to perform analyses rather than to provide a specific analysis. We have not performed a comprehensive evaluation of analysis methods, but offer some suggestions based on a comparison in \cite{KumarThesis}. Two methods that support evaluating the attributes of cost, probability, and time are stochastic-model checking \cite{AFTrees} and game-theoretic analysis \cite{ADTreeValue}. Whichever methods are used must appropriately capture the constraints of our model (including SAND gates) and should be automated to enable rapid attribute-based queries for security metrics such as; the expected attack pay-off for the most likely attack,  or the possible attack pathways given a budget of \$10,000.   

Our approach to constructing the risk model is centered on assets since these are clearly distinguished through Bitcoin descriptors, as continuity of a custodial process, or as operational privacy. However, when performing the risk analysis it can be insightful to consider attacker personas \cite{Shostack}:  a crime syndicate; an opportunistic burglar; a nation state; a business competitor; or even an insider. If the organisation understands any of these personas well (arguably they should especially understand their competitors and employees) they can reduce the uncertainty in their aggregate risk estimates for these scenarios. Attacker-profiles are a useful way to prune attack-trees \cite{KumarThesis}.

\section{Risk Model}
\label{sec:Threat Model}

We have constructed the risk model with several assumptions that limit the scope of the analysis to the operational aspects of custody. Known risks from other protocol and environment dependencies that are discussed in other works should be considered as complementary but are, for the purpose of clarity, assumed to be benign here. First, we assume that the Bitcoin network is functional, realising its live-ness and availability properties \cite{Garay2015,Garay2017,Pass2017,Badertscher2017,Badertscher2018}. We assume that there is significant hash-rate to prevent blockchain reorganisations of a depth higher than the Unvault Tx's relative lock-time. Next, we assume that Revault's Tx model is robust; with scripts that realise the access control structures we expect, without unintended consequences from Tx malleability and network propagation issues as described in \cite{revault-pdf}. We assume the initialization process was secure and safe; private keys and  backups were correctly and confidentially constructed for each participant, software and hardware integrity were verified, relevant public key information for both the wallet and communication was shared among participants leading to a correct configuration for the wallet clients, watchtowers, anti-replay oracles and the coordinator. We assume that Revault's communication security model as described in \cite{practical-revault} is robust. That is, where messages need to be authenticated or confidential, they are. We assume that the software development life-cycle of Revault is secure, such that any deployment is using an implementation that adheres to the protocol specification. Finally, we assume that entities constructing Deposit Txs don't succumb to a man-in-the-middle attack. That is, they lock funds to the deposit descriptor rather than to an attacker's address.

\subsection{Common Attack Sub-Trees}

These attack sub-trees are common to different attacks on Revault, and \textbf{a}, \textbf{b}, \textbf{c}, \textbf{d}, \textbf{e}, \textbf{f} and \textbf{g} are likely to be common to attacks on other custody protocols. 

{\footnotesize
\begin{itemize}[noitemsep,parsep=0pt,partopsep=0pt, leftmargin=0.7cm]
\item[\textbf{a} :] \textbf{Compromise a participant (stakeholder or manager)}
\begin{itemize}[noitemsep,topsep=0pt,parsep=0pt,partopsep=0pt, leftmargin=0.8cm]
\item[1 :] Coerce participant (OR)
\item[2 :] Corrupt participant
\end{itemize}
\end{itemize}
}

\noindent Coercion and insider threats from corrupt participants must be considered. Legal defences for malicious employee behaviour can be effective deterrents here. 

{\footnotesize
\begin{itemize}[noitemsep,parsep=0pt,partopsep=0pt, leftmargin=0.7cm]
\item[\textbf{b} :] \textbf{Compromise a participant's (stakeholder's or manager's) HSM }
\begin{itemize}[noitemsep,topsep=0pt,parsep=0pt,partopsep=0pt, leftmargin=0.8cm]
\item[1 :] Physical attack of HSM (OR)
\begin{itemize}[noitemsep,topsep=0pt,parsep=0pt,partopsep=0pt, leftmargin=0.9cm]
\item[\textit{1.1} :] Determine location of participant’s HSM (SAND)
\item[\textit{1.2} :] Access the physical security environment of the participant’s HSM (SAND)
\item[\textit{1.3} :] Exfiltrate keys (either on premise or after stealing it) (OR)
\item[\textit{1.4} :] By-pass PIN and make the HSM sign a malicious chosen message
\end{itemize}
\item[2 :] Remote attack of HSM (OR)
\begin{itemize}[noitemsep,topsep=0pt,parsep=0pt,partopsep=0pt, leftmargin=0.9cm]
\item[\textit{2.1} :] Compromise a device that is then connected to the HSM (SAND)
\begin{itemize}[noitemsep,topsep=0pt,parsep=0pt,partopsep=0pt, leftmargin=1cm]
\item[\textit{2.1.1} :] (see \textbf{g}) Compromise the participant's wallet software (OR)
\item[\textit{2.1.2} :] Trick participant into connecting their HSM to a compromised device via social engineering
\end{itemize}
\item[\textit{2.2} :] Exploit a firmware vulnerability (OR)
\item[\textit{2.3} :] Trick participant into compromising their own HSM with the user interface of the compromised device
\end{itemize}
\item[3 :](see \textbf{a})  Compromise a participant
\end{itemize}
\end{itemize}
}

{\footnotesize
\begin{itemize}[noitemsep,parsep=0pt,partopsep=0pt, leftmargin=0.7cm]
\item[\textbf{c} :] \textbf{Compromise a participant’s (stakeholder's or manager's) keys backup}
\begin{itemize}[noitemsep,topsep=0pt,parsep=0pt,partopsep=0pt, leftmargin=0.9cm]
\item[1 :] Physical Attack (OR)
\begin{itemize}[noitemsep,topsep=0pt,parsep=0pt,partopsep=0pt, leftmargin=0.9cm]
\item[\textit{1.1} :] Determine location of the keys backup (SAND)
\begin{itemize}[noitemsep,topsep=0pt,parsep=0pt,partopsep=0pt, leftmargin=1cm]
\item[\textit{1.1.1} :] Watch the participant between the custody initialization and the start of operations (OR)
\item[\textit{1.1.2} :] Watch the participant during a backup check (OR)
\end{itemize}
\item[\textit{1.2} :] Access the physical security environment of the keys backup (SAND)
\item[\textit{1.3} :] Depending on backup format, steal or copy it
\end{itemize}
\item[2 :] (see \textbf{a}) Compromise a participant
\end{itemize}
\end{itemize}
}

{\footnotesize
\begin{itemize}[noitemsep,parsep=0pt,partopsep=0pt, leftmargin=0.7cm]
\item[\textbf{d} :] \textbf{Compromise a server (watchtower, anti-replay oracle, or coordinator)}
\begin{itemize}[noitemsep,topsep=0pt,parsep=0pt,partopsep=0pt, leftmargin=0.8cm]
\item[1 :] Remote attack (OR)
\begin{itemize}[noitemsep,topsep=0pt,parsep=0pt,partopsep=0pt, leftmargin=0.9cm]
\item[\textit{1.1} :] Exploit a software vulnerability (OR)
\begin{itemize}[noitemsep,topsep=0pt,parsep=0pt,partopsep=0pt, leftmargin=0.9cm]
\item[\textit{1.1.1} :] Determine the public interfaces of the server (SAND)
\item[\textit{1.1.2} :] Exploit a vulnerability on one of the softwares listening on these interfaces
\end{itemize}
\item[\textit{1.2} :] Exploit a human vulnerability (e.g. trick participant into performing a malicious update)
\end{itemize}
\item[2 :] Physical attack (OR)
\begin{itemize}[noitemsep,topsep=0pt,parsep=0pt,partopsep=0pt, leftmargin=0.9cm]
\item[\textit{2.1} :] Determine server's location (SAND)
\item[\textit{2.2} :] Access the physical security environment of the server (SAND)
\end{itemize}
\item[3 :] (see \textbf{a}) Compromise the participant managing the server
\end{itemize}
\end{itemize}
}

\noindent An attacker who successfully completes \textbf{d} for a watchtower will be able to steal funds from the watchtower's fee wallet and will be able to force an emergency scenario by broadcasting all Emergency and Unvault-Emergency Txs it has stored. They can also prevent broadcast of a Cancel Tx from this watchtower either passively (\textit{ACK} the secure storage of the signature to the stakeholder, but then drop the signature) or actively.

{\footnotesize
\begin{itemize}[noitemsep,parsep=0pt,partopsep=0pt, leftmargin=0.7cm]
\item[\textbf{e} :] \textbf{Shutdown a watchtower}
\begin{itemize}[noitemsep,topsep=0pt,parsep=0pt,partopsep=0pt, leftmargin=0.8cm]
\item[1 :] Physical attack on the watchtower (OR)
\begin{itemize}[noitemsep,topsep=0pt,parsep=0pt,partopsep=0pt, leftmargin=0.9cm]
\item[\textit{1.1} :] Determine watchtower's location (SAND)
\item[\textit{1.2} :] Sever the internet connection to the building in which the watchtower is located (OR)
\item[\textit{1.3} :] Sever the power-line connection to the building in which the watchtower is located (OR)
\item[\textit{1.4} :] Access the physical security of the watchtower and un-plug the machine
\end{itemize}
\item[2 :] Remote attack on the watchtower
\begin{itemize}[noitemsep,topsep=0pt,parsep=0pt,partopsep=0pt, leftmargin=0.9cm]
\item[\textit{2.1} :] Determine public interfaces of watchtower (SAND)
\item[\textit{2.2} :] Denial of Service attack through one of the public interfaces (OR)
\item[\textit{2.3} :] Eclipse attack on the watchtower's Bitcoin node \cite{EclipseAttack} (OR)
\begin{itemize}[noitemsep,topsep=0pt,parsep=0pt,partopsep=0pt, leftmargin=1cm]
\item[\textit{2.3.1} :] Slowly force de-synchronisation of watchtower with the true block height by delaying block propagation \cite{TimeDilationAttack} (OR)
\item[\textit{2.3.2} :] Prevent outgoing propagation of Cancel or Emergency Txs
\end{itemize}
\item[\textit{2.4} :] Denial of Service attack on the fee-bumping UTxOs pool---not enough funds to pay competitive fees (OR)
\end{itemize}
\end{itemize}
\end{itemize}
}

{\footnotesize
\begin{itemize}[noitemsep,parsep=0pt,partopsep=0pt, leftmargin=0.7cm]
\item[\textbf{f} :] \textbf{Get signature from participant to unlock UTxO for Theft Tx}
\begin{itemize}[noitemsep,topsep=0pt,parsep=0pt,partopsep=0pt, leftmargin=0.8cm]
\item[1 :] (see \textbf{a}) Compromise a participant (OR)
\item[2 :] (see \textbf{b}) Compromise participant's HSM (OR)
\item[3 :] (see \textbf{c}) Compromise participant's keys backup
\end{itemize}
\end{itemize}
}

{\footnotesize
\begin{itemize}[noitemsep,parsep=0pt,partopsep=0pt, leftmargin=0.7cm]
\item[\textbf{g} :] \textbf{Compromise a participant's wallet}
\begin{itemize}[noitemsep,topsep=0pt,parsep=0pt,partopsep=0pt, leftmargin=0.8cm]
\item[1 :] Physical attack (OR)
\begin{itemize}[noitemsep,topsep=0pt,parsep=0pt,partopsep=0pt, leftmargin=0.9cm]
\item[\textit{1.1} :] Locate participant's device (SAND)
\item[\textit{1.2} :] Access physical security environment of participant's device
\end{itemize}
\item[2 :] Remote attack (OR)
\begin{itemize}[noitemsep,topsep=0pt,parsep=0pt,partopsep=0pt, leftmargin=0.9cm]
\item[\textit{2.1} :] Determine public interfaces of device (SAND)
\item[\textit{2.2} :] Exploit a vulnerability
\end{itemize}
\item[3 :] (see \textbf{a}) Compromise participant
\end{itemize}
\end{itemize}
}

\noindent Participant's wallet devices are expected to be used for day-to-day activities. With many vulnerabilities to exploit, the likelihood of success for \textbf{g} is high.

{\footnotesize
\begin{itemize}[noitemsep,parsep=0pt,partopsep=0pt, leftmargin=0.7cm]
\item[\textbf{h} :] \textbf{Determine the locking Script for a deposit or unvault UTxO (\textit{Witness Script})}
\begin{itemize}[noitemsep,topsep=0pt,parsep=0pt,partopsep=0pt, leftmargin=0.8cm]
\item[1 :] (see \textbf{g}) Compromise any participant's wallet (OR)
\item[2 :] (see \textbf{d}) Compromise a watchtower (OR)
\item[3 :] (see \textbf{d}) Compromise an anti-replay oracle
\end{itemize}
\end{itemize}
}

\noindent Deposit and unvault descriptors are deterministic, but public keys are needed to derive UTxO locking Scripts. These are stored by all wallets, watchtowers and anti-replay oracles.

{\footnotesize
\begin{itemize}[noitemsep,parsep=0pt,partopsep=0pt, leftmargin=0.7cm]
\item[\textbf{i} :] \textbf{Satisfy an input in a Theft Tx that consumes an identified deposit UTxO or unvault UTxO (through $N-$of$-N$)}
\begin{itemize}[noitemsep,topsep=0pt,parsep=0pt,partopsep=0pt, leftmargin=0.8cm]
\item[1 :] (see \textbf{h}) Determine the UTxO locking Script (\textit{Witness Script}) (SAND)
\item[2 :] Prevent the relevant Emergency Tx from being broadcast until the Theft Tx is confirmed (where $A + B = N$) (AND)
\begin{itemize}[noitemsep,topsep=0pt,parsep=0pt,partopsep=0pt, leftmargin=0.9cm]
\item[\textit{2.1} :] (see \textbf{d}) Compromise a watchtower (A times)
\item[\textit{2.2} :] (see \textbf{e}) Shutdown a watchtower (B times)
\item[\textit{2.3} :] (see \textbf{g}) Compromise stakeholder's wallet ($N$ times)
\end{itemize}
\item[3 :] (see \textbf{f}) Get signature from a stakeholder to unlock UTxO for Theft Tx ($N$ times)
\end{itemize}
\end{itemize}
}

{\footnotesize
\begin{itemize}[noitemsep,parsep=0pt,partopsep=0pt, leftmargin=0.7cm]
\item[\textbf{j} :] \textbf{Satisfy an input in a Theft Tx that consumes an identified unvault UTxO (through $K-$of$-M$, anti-replay oracles and time-lock)}
\begin{itemize}[noitemsep,topsep=0pt,parsep=0pt,partopsep=0pt, leftmargin=0.8cm]
\item[1 :] (see \textbf{h}) Determine the UTxO locking Script (\textit{Witness Script}) (SAND)
\item[2 :] Receive signatures for Theft Tx from all $N$ anti-replay oracles (AND)
\begin{itemize}[noitemsep,topsep=0pt,parsep=0pt,partopsep=0pt, leftmargin=0.9cm]
\item[\textit{2.1} :] Compromise a manager’s private communication keys and the set of anti-replay oracles’ public communication keys (OR)
\begin{itemize}[noitemsep,topsep=0pt,parsep=0pt,partopsep=0pt, leftmargin=1cm]
\item[\textit{2.1.1} :] (see \textbf{g}) Compromise a manager’s wallet (OR)
\item[\textit{2.1.2} :] (see \textbf{a}) Compromise a manager
\end{itemize}
\item[\textit{2.2} :] (see \textbf{d}) Compromise the anti-replay oracle
\end{itemize}
\item[3 :] (see \textbf{f}) Get signature from a manager to unlock UTxO for Theft Tx ($K$ times)
\end{itemize}
\end{itemize}
}

{\footnotesize
\begin{itemize}[noitemsep,parsep=0pt,partopsep=0pt, leftmargin=0.7cm]
\item[\textbf{k} :] \textbf{Satisfy an input in a Theft Tx that consumes an identified emergency UTxO}
\begin{itemize}[noitemsep,topsep=0pt,parsep=0pt,partopsep=0pt, leftmargin=0.8cm]
\item[1 :] Determine the emergency descriptor policy (SAND)
\item[2 :] Satisfy the emergency descriptor's locking conditions (may include waiting for time-locks, giving sufficient signatures, giving hash pre-images, \textit{etc}.)
\end{itemize}
\end{itemize}
}

\noindent The details of the emergency descriptor are intentionally not specified with the Revault protocol, except that it is more difficult to access than the deposit descriptor. Stakeholders may compartmentalise and distribute the descriptor information to afford its privacy some resilience to attack.

\subsection{Attack-Trees}
\label{subsec:attck-trees}

The following attack-trees are the foundation for an operational risk framework for Revault.

{\footnotesize
\begin{itemize}[noitemsep,parsep=0pt,partopsep=0pt, leftmargin=0.7cm]
\item[\textbf{A} :] \textbf{Compromise privacy of the custody operation (determine the set of public UTxOs)}
\begin{itemize}[noitemsep,topsep=0pt,parsep=0pt,partopsep=0pt, leftmargin=0.8cm]
\item[1 :] (see \textbf{d}) Compromise any of the servers (OR)
\item[2 :] (see \textbf{a}) Compromise a participant (OR)
\item[3 :] (see \textbf{g}) Compromise a participant's wallet (OR)
\item[4 :] Traffic analysis of connections between servers and/or wallets (OR)
\item[5 :] Blockchain analysis
\end{itemize}
\end{itemize}
}

\noindent Without privacy support for advanced descriptors (such as by using MuSig2 \cite{MuSig2} or MuSig-DN \cite{MuSig-DN} if the proposed Taproot \cite{Taproot} upgrade is activated by the Bitcoin network) Revault's operational privacy is brittle. 

{\footnotesize
\begin{itemize}[noitemsep,parsep=0pt,partopsep=0pt, leftmargin=0.7cm]
\item[\textbf{B} :] \textbf{Broadcast Theft Tx(s) that consume all deposit UTxOs}
\begin{itemize}[noitemsep,topsep=0pt,parsep=0pt,partopsep=0pt, leftmargin=0.8cm]
\item[1 :] (see \textbf{A}) Determine $\mathcal{D}$, the set of deposit UTxOs (SAND)
\item[2 :] (see \textbf{h}) Determine the locking Script for deposit UTxO ($|\mathcal{D}|$ times)
\item[3 :] (see \textbf{i}) Satisfy an input in a Theft Tx that consumes an identified deposit UTxO ($|\mathcal{D}|$ times) 
\end{itemize}
\end{itemize}
}

\noindent A Theft Tx that consumes all available deposit UTxOs would be catastrophic since this comprises the majority of funds. We recommend a defence wherein each stakeholder is equipped with a panic button that is directly connected to their watchtower or dedicated emergency service. When triggered, all the signed Emergency and Unvault-Emergency Txs are broadcast, negating the pay-off for an attacker and thus acting as a deterrent. 

{\footnotesize
\begin{itemize}[noitemsep,parsep=0pt,partopsep=0pt, leftmargin=0.7cm]
\item[\textbf{C} :] \textbf{Broadcast Theft Tx(s) that consume as many available unvault UTxOs as watchtower spending policies permit}
\begin{itemize}[noitemsep,topsep=0pt,parsep=0pt,partopsep=0pt, leftmargin=0.8cm]
\item[1 :] Determine spending constraints of all watchtowers' policies (SAND)
\begin{itemize}[noitemsep,topsep=0pt,parsep=0pt,partopsep=0pt, leftmargin=0.9cm]
\item[\textit{1.1} :] (see \textbf{a}) Compromise a participant (OR)
\item[\textit{1.2} :] (see \textbf{g}) Compromise a manager's wallet 
\item[\textit{1.3} :] (see \textbf{d}) Compromise a watchtower ($N$ times)
\end{itemize}
\item[2 :] Determine $\mathcal{U}$, the set of available unvault UTxOs (SAND)
\begin{itemize}[noitemsep,topsep=0pt,parsep=0pt,partopsep=0pt, leftmargin=0.9cm]
\item[\textit{2.1} :] (see \textbf{A}) Compromise privacy of the custody operation (determine the set of public UTxOs) (SAND)
\item[\textit{2.2} :] (see \textbf{h}) Determine the locking Script for unvault UTxO ($|\mathcal{U}|$ times)
\end{itemize}
\item[3 :] (see \textbf{i} OR \textbf{j}) Satisfy  an  input  in  a  Theft  Tx  that  consumes  an  identified  unvault UTxO ($|\mathcal{U}|$ times)
\end{itemize}
\end{itemize}
}

\noindent \textbf{C} can be avoided if watchtowers have a white-list of addresses that Spend Txs can pay to.

{\footnotesize
\begin{itemize}[noitemsep,parsep=0pt,partopsep=0pt, leftmargin=0.7cm]
\item[\textbf{D} :] \textbf{Broadcast Theft Tx(s) that consume all available unvault UTxOs, by-passing watchtowers' spending policies}
\begin{itemize}[noitemsep,topsep=0pt,parsep=0pt,partopsep=0pt, leftmargin=0.8cm]
\item[1 :] Prevent watchtower from broadcasting Cancel or Unvault-Emergency Txs before Theft Tx is confirmed ($N$ times SAND)
\begin{itemize}[noitemsep,topsep=0pt,parsep=0pt,partopsep=0pt, leftmargin=0.9cm]
\item[\textit{1.1} :] (see \textbf{d}) Compromise a watchtower (OR)
\item[\textit{1.2} :] (see \textbf{e}) Shutdown a watchtower
\end{itemize}
\item[2 :] Determine $\mathcal{U}$, the set of available unvault UTxOs (SAND)
\begin{itemize}[noitemsep,topsep=0pt,parsep=0pt,partopsep=0pt, leftmargin=0.9cm]
\item[\textit{2.1} :] (see \textbf{A}) Compromise privacy of the custody operation (determine the set of public UTxOs) (SAND)
\item[\textit{2.2} :] (see \textbf{h}) Determine the locking Script for unvault UTxO ($|\mathcal{U}|$ times)
\end{itemize}
\item[3 :] (see \textbf{i} OR \textbf{j}) Satisfy  an  input  in  a  Theft  Tx  that  consumes  an  identified  unvault UTxO ($|\mathcal{U}|$ times)
\end{itemize}
\end{itemize}
}

{\footnotesize
\begin{itemize}[noitemsep,parsep=0pt,partopsep=0pt, leftmargin=0.7cm]
\item[\textbf{E} :] \textbf{Broadcast a Theft Tx that by-passes watchtowers' spending policies}
\begin{itemize}[noitemsep,topsep=0pt,parsep=0pt,partopsep=0pt, leftmargin=0.8cm]
\item[1 :] Determine $\mathcal{U}$, the set of available unvault UTxOs (SAND)
\begin{itemize}[noitemsep,topsep=0pt,parsep=0pt,partopsep=0pt, leftmargin=0.9cm]
\item[\textit{1.1} :] (see \textbf{A}) Compromise privacy of the custody operation (determine the set of public UTxOs) (SAND)
\item[\textit{1.2} :] (see \textbf{h}) Determine the locking Script for unvault UTxO ($|\mathcal{U}|$ times)
\end{itemize}
\item[2 :] (see \textbf{f}) Get signature from a manager to unlock $U \subseteq \mathcal{U}$, a subset of available unvault UTxOs for a valid Spend Tx ($K$ times)
\item[3 :] (see \textbf{i} OR \textbf{j}) Satisfy  an  input  in  a  Theft  Tx  that  consumes  an  identified  unvault UTxO ($|U|$ times)
\item[4 :] (see \textbf{d}) Compromise an anti-replay oracle to get a signature for the valid Spend Tx which consumes $U$, the UTxOs ($N$ times SAND)
\item[5 :] Advertise the valid Spend Tx to the watchtowers through the coordinator (SAND)
\item[6 :] Broadcast all Unvault Txs that the valid Spend Tx depends on and wait for the time-lock to expire
\end{itemize}
\end{itemize}
}

{\footnotesize
\begin{itemize}[noitemsep,parsep=0pt,partopsep=0pt, leftmargin=0.7cm]
\item[\textbf{F} :] \textbf{Force emergency scenario}
\begin{itemize}[noitemsep,topsep=0pt,parsep=0pt,partopsep=0pt, leftmargin=0.8cm]
\item[1 :] Broadcast the full set of signed Emergency and Unvault-emergency transactions 
\begin{itemize}[noitemsep,topsep=0pt,parsep=0pt,partopsep=0pt, leftmargin=0.9cm]
\item[\textit{1.1} :] (see \textbf{d}) Compromise a watchtower (OR)
\item[\textit{1.2} :] (see \textbf{a}) Compromise a stakeholder
\end{itemize}
\end{itemize}
\end{itemize}
}

\noindent The emergency deterrent results in better security from the most egregious physical threats to participants (particularly stakeholders who control the majority of funds) but also in a fragility to the continuity of operations that could be abused by an attacker. Attacks that rely on \textbf{E} may seek a pay-off other than fund theft, such as damaging the reputation of the organisation for having down-time and taking a leveraged bet on the likely market consequences. However, forced down-time attacks through power or internet outages or detainment of personnel are prevalent threats for organisations who aren't deploying Revault. In any case, with this risk model the consequence of not using an emergency deterrent can be considered by performing an analysis with pruned attack-trees.

{\footnotesize
\begin{itemize}[noitemsep,parsep=0pt,partopsep=0pt, leftmargin=0.7cm]
\item[\textbf{G} :] \textbf{Broadcast a Theft Tx which consumes all available UTxOs locked to the emergency descriptor}
\begin{itemize}[noitemsep,topsep=0pt,parsep=0pt,partopsep=0pt, leftmargin=0.8cm]
\item[1 :] (see \textbf{F}) Force an emergency scenario (SAND)
\item[2 :] Determine $\mathcal{E}$, the set of available emergency UTxOs (SAND)
\begin{itemize}[noitemsep,topsep=0pt,parsep=0pt,partopsep=0pt, leftmargin=0.9cm]
\item[\textit{2.1} :] (see \textbf{A}) Compromise privacy of the custody operation (determine the set of public UTxOs)
\end{itemize}
\item[3 :] (see \textbf{k}) Satisfy an input in a Theft Tx that consumes an identified emergency UTxO ($|\mathcal{E}|$ times)
\end{itemize}
\end{itemize}
}

{\footnotesize
\begin{itemize}[noitemsep,parsep=0pt,partopsep=0pt, leftmargin=0.7cm]
\item[\textbf{H} :] \textbf{Broadcast a Theft Tx which spends from a manager’s fee wallet}
\begin{itemize}[noitemsep,topsep=0pt,parsep=0pt,partopsep=0pt, leftmargin=0.8cm]
\item[1 :] (see \textbf{g}) Compromise a manager’s wallet
\end{itemize}
\end{itemize}
}

\noindent While this is a relatively simple attack, the fee wallet will never hold a significant portion of bitcoins and is considered external to the custody protocol.

{\footnotesize
\begin{itemize}[noitemsep,parsep=0pt,partopsep=0pt, leftmargin=0.7cm]
\item[\textbf{I} :] \textbf{Prevent Emergency, Unvault-Emergency, and Cancel Tx valid signature exchange}
\begin{itemize}[noitemsep,topsep=0pt,parsep=0pt,partopsep=0pt, leftmargin=0.8cm]
\item[1 :] $1$ of $N$ stakeholders doesn't sign (OR)
\begin{itemize}[noitemsep,topsep=0pt,parsep=0pt,partopsep=0pt, leftmargin=0.9cm]
\item[\textit{1.1} :] Prevent stakeholder from accessing their HSM (OR)
\item[\textit{1.2} :] Prevent stakeholder from accessing their wallet (OR)
\item[\textit{1.3} :] (see \textbf{a}) Compromise a stakeholder
\end{itemize}
\item[2 :] Shutdown coordinator (OR)
\item[3 :] (see \textbf{e}) Shutdown a watchtower ($N$ times) (OR)
\item[4 :] Blockchain re-organization and Deposit Tx outpoint malleation.
\end{itemize}
\end{itemize}
}

{\footnotesize
\begin{itemize}[noitemsep,parsep=0pt,partopsep=0pt, leftmargin=0.7cm]
\item[\textbf{J} :] \textbf{Prevent Unvault Tx signature exchange}
\begin{itemize}[noitemsep,topsep=0pt,parsep=0pt,partopsep=0pt, leftmargin=0.8cm]
\item[1 :] $1$ of $N$ stakeholders doesn't sign (OR)
\begin{itemize}[noitemsep,topsep=0pt,parsep=0pt,partopsep=0pt, leftmargin=0.9cm]
\item[\textit{1.1} :] Prevent stakeholder from accessing their HSM (OR)
\item[\textit{1.2} :] Prevent stakeholder from accessing their wallet software (OR)
\item[\textit{1.3} :] (see \textbf{a}) Compromise a stakeholder
\end{itemize}
\item[2 :] Shutdown coordinator (OR)
\item[3 :] Prevent all managers from accessing their wallet software
\end{itemize}
\end{itemize}
}

{\footnotesize
\begin{itemize}[noitemsep,parsep=0pt,partopsep=0pt, leftmargin=0.7cm]
\item[\textbf{K} :] \textbf{Prevent managers from broadcasting a Spend Tx}
\begin{itemize}[noitemsep,topsep=0pt,parsep=0pt,partopsep=0pt, leftmargin=0.8cm]
\item[1 :] Prevent managers from signing the Spend transaction (OR)
\begin{itemize}[noitemsep,topsep=0pt,parsep=0pt,partopsep=0pt, leftmargin=0.9cm]
\item[\textit{1.1} :] (see \textbf{d}) Compromise an anti-replay oracle (OR)
\item[\textit{1.2} :] Prevent sufficient threshold of managers from signing the Spend Tx
(where $A + B + C = M-K+1$) (OR) 
\begin{itemize}[noitemsep,topsep=0pt,parsep=0pt,partopsep=0pt, leftmargin=1cm]
\item[\textit{1.2.1} :] (see \textbf{a}) Compromise a manager ($A$ times)
\item[\textit{1.2.2} :] Prevent manager from accessing their HSM ($B$ times)
\item[\textit{1.2.3} :] Prevent manager from accessing their wallet software ($C$ times)
\end{itemize}
\end{itemize}
\item[2 :] Force broadcast of Cancel Tx (OR)
\begin{itemize}[noitemsep,topsep=0pt,parsep=0pt,partopsep=0pt, leftmargin=0.9cm]
\item[\textit{2.1} :] (see \textbf{d}) Compromise a watchtower 
\end{itemize}
\item[3 :] Prevent broadcast of Unvault Tx
\begin{itemize}[noitemsep,topsep=0pt,parsep=0pt,partopsep=0pt, leftmargin=0.8cm]
\item[\textit{3.1} :] High demand for block space making the Unvault Tx not profitable to mine.\footnote{Manager's fee-bumping wallet can not cover this until a network policy such as Package Relay \cite{PackageRelay} is implemented.} %
\item[\textit{3.2} :] (see \textbf{g}) Compromise manager's wallet ($M$ times)
\end{itemize}
\end{itemize}
\end{itemize}
 }

\section{Conclusion}
\label{sec:Conclusion}

The rise of Bitcoin has led to a new commercial ecosystem, with market exchanges enabling its sale and purchase, companies and financial institutions offering secure custody services, and insurance brokers and underwriters willing to insure individuals, exchanges and custodians against loss or theft of their assets
. In this paper we first posit that a methodology to better understand risks in custodial operations is needed, something complementary to understanding blockchain and cryptographic security. We put forth requirements of the modelling technique and propose attack-trees as a formalism which satisfies those requirements. We exemplify the approach by presenting a library of attack-trees constructed for a multi-party custody protocol called Revault and explain how this framework can be used as a basis for risk-management in custodial operations. The next steps for this work are to: construct a set of defences to the prominent risks and incorporate them into the model; and to determine or build a suitable tool for automating computations for a specific analysis. 

\section*{Acknowledgements}

We thank Professor McBurney (King's College London), Kevin Laoec (WizardSardine) for insightful conversations and for reviewing the text. 

\section*{Funding}

Funding is gratefully acknowledged under a UK EPSRC-funded GTA Award through King's College London and from WizardSardine.

\bibliographystyle{splncs04}
\bibliography{sample}

\end{document}